\documentclass[aps,pra,amsmath,amssymb,twocolumn,groupedaddress]{revtex4}

\usepackage{graphicx,epsfig} 

\usepackage[ps2pdf,colorlinks]{hyperref}
\usepackage{mathptm}

\newcommand{\ket}[1]{|#1\rangle} \newcommand{\bra}[1]{\langle#1|}

\begin{document}

\title{A toolbox for lattice spin models with polar molecules}
\author{A.~Micheli}\email{andrea.micheli@uibk.ac.at}
\author{G.~K.~Brennen}
\author{P.~Zoller}
\affiliation{Institute for Theoretical Physics,
  University of Innsbruck, and Institute for Quantum Optics and
  Quantum Information of the Austrian Academy of Science, 6020
  Innsbruck, Austria} \date{\today}

\begin{abstract}
There is growing interest to investigate states of matter
with topological order, which support excitations in the form of
anyons, and which underly topological quantum computing. Examples of
such systems include lattice spin models in two dimensions.
Here we show that relevant Hamiltonians can be systematically
engineered with polar molecules stored in optical lattices, where
the spin is represented by a single electron outside a closed shell
of a heteronuclear molecule in its rotational ground state.
Combining microwave excitation with the dipole-dipole interactions
and spin-rotation couplings allows us to build a complete toolbox
for effective two-spin interactions with designable range and
spatial anisotropy, and with coupling strengths significantly larger
than relevant decoherence rates. As an illustration we discuss two
models: a $2$D square lattice with an energy gap providing for
protected quantum memory, and another on stacked triangular lattices
leading to topological quantum computing.
\end{abstract}


\maketitle

\section{Introduction}

Lattice spin models are ubiquitous in condensed matter physics where
they are used as simplified models to describe the characteristic
behavior of more complicated interacting physical systems. Recently
there have been exciting theoretical discoveries of models with
quasi-local spin interactions with emergent topological order
\cite{Wen:03,Hermele:04}.  In contrast to Landau theory where
various phases of matter are described by broken symmetries,
topological ordered states are distinguished by homology class and
have the property of being robust to arbitrary perturbations of the
underlying Hamiltonian.  These states do not exhibit long range
order in pairwise operators, rather they have long range order in
highly nonlocal strings of operators.  A real world example is the
fractional quantum Hall effect which gives rise to states with the
same symmetry but distinguishable by quantum numbers associated with
the topology of the surface they live on \cite{Einarsson}.

It is of significant interest to ``design'' materials with these
properties, both to observe and study exotic phases, and in light of
possible applications. Cold atomic and molecular gases in optical
lattices are prime candidates for this endeavor in view of the
complete controllability of these systems in the laboratory. The
idea of realizing bosonic and fermionic Hubbard models, and thus
also lattice spin models, with cold atoms in optical lattices has
sparked a remarkable series of experiments, and has triggered
numerous theoretical studies to develop cold atoms as a quantum
simulator for strongly correlated condensed matter systems (see
e.g. \cite{Jaksch, Buechler:04,Santos:04} and references therein). However, coaxing a
physical system to mimic the required interactions for relevant
lattice spin models, which must be both anisotropic in space and in
the spin degrees of freedom, and given range, is highly nontrivial.
Here we show that cold gases of polar molecules, as presently
developed in the laboratory \cite{PolMol}, allow us to construct in
a natural way a {\em complete toolbox} for any permutation symmetric
two spin-$1/2$ (qubit) interaction. The attractiveness of this idea
also rests on the fact that dipolar interactions have coupling
strengths significantly larger than those of the atomic Hubbard
models, and relevant decoherence rates.

Our basic building block is a system of two polar molecules strongly
trapped a given sites of an optical lattice, where the spin-$1/2$
(or qubit) is represented by a single electron outside a closed
shell of a heteronuclear molecule in its rotational ground state.
Heteronuclear molecules have large permanent electric dipole
moments. This implies that the rotational motion of molecules is
coupled {\em strongly} via the dipole-dipole interactions, whose
signatures are the {\em long range} $1/r^3$ character and an {\em
angular dependence}, where the polar molecules attract or repel each
other depending on the relative orientation of their dipole moments.
In addition, microwave excitation of rotational energy levels allows
to effectively tailor the \emph{spatial dependence} of dipole-dipole
interactions. Finally, accounting for the spin-rotation splitting of
molecular rotational levels we can make these dipole-dipole
interactions \emph{spin-dependent}. General lattice spin models are
readily built from these binary interactions.

\section{ANISOTROPIC SPIN MODELS WITH NOISE RESILIENT GROUND STATES}
\begin{figure}[htb]
  \begin{center}
    \includegraphics[width=.99\columnwidth]{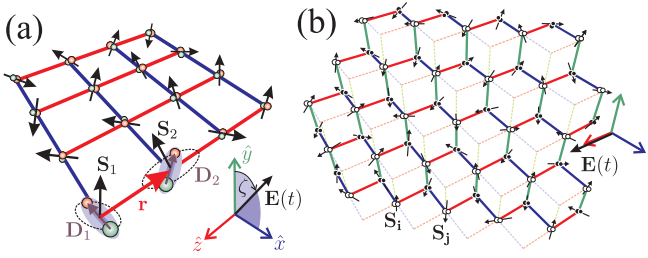}
    \caption{\label{fig:1}(Color online) Example anisotropic spin models
    that can be simulated with polar molecules trapped in optical lattices.
   (a) Square lattice in $2$D with nearest neighbor orientation dependent Ising interactions
   along $\hat{x}$ and $\hat{z}$.  Effective interactions between the spins
      ${\bf S}_1$ and ${\bf S}_2$ of the molecules in their rovibrational ground states
      are generated with a
      microwave field ${\bf E}(t)$ inducing
      dipole-dipole interactions between the molecules
      with dipole moments ${\bf D}_1$ and ${\bf D}_2$, respectively.
     (b) Two staggered triangular lattices with nearest neighbors oriented along
     orthogonal triads. The interactions depend on the orientation of the links with
     respect to the electric field.  (Dashed lines are included for perspective.)
    }
  \end{center}
\end{figure}

  Two highly anisotropic models with spin-$1/2$
  particles which we will show
  how to simulate are illustrated in Figs.~
  \ref{fig:1}a and \ref{fig:1}b respectively.  The first takes place on a square $2$D lattice with nearest
  neighbor interactions
\begin{equation}
  H_{\rm spin}^{({\rm I})}= \sum_{i=1}^{\ell -1}\sum_{j=1}^{\ell -1} J (\sigma^z_{i,j}\sigma^z_{i,j+1}+\cos\zeta\sigma^x_{i,j}\sigma^x_{i+1,j}).
  \label{Ioffe}
\end{equation}
Introduced by Duo\c{c}ot {\it et al.} \cite{Duocot:05} in the context of 
Josephson junction arrays, this model
(for $\zeta\neq \pm\pi/2$) admits a 2- fold degenerate ground
subspace that is immune to local noise up to $\ell$th order and
hence is a good candidate for storing a protected qubit.

The second, occurs on a bipartite lattice constructed with two
$2$D triangular lattices, one shifted and stacked on top of the
other.  The interactions are indicated by nearest neighbor links
along the $\hat{x}, \hat{y}$ and $\hat{z}$ directions in real space:
\begin{equation}
H_{\rm spin}^{({\rm II})}=J_{\perp}\sum_{x-{\rm
links}}\sigma^x_j\sigma^x_k+ J_{\perp} \sum_{y-{\rm
links}}\sigma^y_j\sigma^y_k+ J_z\sum_{z-{\rm
links}}\sigma^z_j\sigma^z_k. \label{Kit}
\end{equation}
This model has the same spin dependence and nearest neighbor graph
as the model on a honeycomb lattice introduced by Kitaev
\cite{Kitaev:05}. He has shown that by tuning the ratio of
interaction strengths $|J_{\perp}|/|J_z|$ one can tune the system
from a gapped phase carrying Abelian anyonic excitations to a
gapless phase which in the presence of a magnetic field becomes
gapped with non-Abelian excitations. In the regime
$|J_{\perp}|/|J_z|\ll 1$ the Hamilonian can be mapped to a model
with four body operators on a square lattice with ground states that
encode topologically protected quantum memory \cite{DKL:03}. One
proposal \cite{Duan:03} describes how to use trapped atoms in
spin dependent optical lattices to simulate the spin model $H_{\rm
spin}^{({\rm II})}$. There the induced spin couplings are obtained 
via spin dependent collisions in second
order tunneling processes.  Larger coupling strengths are desirable. 
In both spin models (${\rm I}$ and ${\rm II}$) above, the signs of the interactions are
irrelevant although we will be able to tune the signs if needed.

\section{SPECTROSCOPY OF POLAR MOLECULES IN OPTICAL LATTICES}



Our system is comprised of heteronuclear molecules with $^{2}
\Sigma_{1/2}$ ground electronic states, corresponding for example to
alkaline-earth monohalogenides with a single electron outside a
closed
shell. We adopt a model molecule 
where the rotational excitations are described by the Hamiltonian
$H_{\rm m} =B \bf N^2 + \gamma \bf N\cdot \bf S$ with $\bf N$ the
dimensionless orbital angular momentum of the nuclei, and $\bf S$ the 
dimensionless electronic
spin (assumed to be $S=1/2$ in the following). Here $B$ denotes the
rotational constant and $\gamma$ is the spin-rotation coupling
constant, where a typical $B$ is a few tens of GHz, and $\gamma$ in
the hundred MHz regime. The coupled basis of a single molecule $i$
corresponding to the eigenbasis of $H_{\rm m}^i$ is
$\{\ket{N_i,S_i,J_i;M_{J_i}}\}$ where ${\bf
  J}_i={\bf N}_i+{\bf S}_i$ with eigenvalues
$E(N=0,1/2,1/2)=0,E(1,1/2,1/2)=2B-\gamma$, and
$E(1,1/2,3/2)=2B+\gamma/2$.  While we ignore hyperfine interactions
in the present work, our discussion below is readily extended to
include hyperfine effects, which offer extensions to spin systems
$S>1/2$.

The Hamiltonian describing the internal and external dynamics of a
pair of molecules trapped in wells of an optical lattice is denoted
by $H=H_{\rm in}+H_{\rm ex}$. The interaction describing the
internal degrees of freedom is $H_{\rm in}=H_{\rm dd}+\sum_{i=1}^2
H_{\rm m}^i$. Here $H_{\rm dd}$ is the dipole-dipole interaction
given below in Eq.~\eqref{dd}. The Hamiltonian describing the
external, or motional, degrees of freedom is $H_{\rm
ex}=\sum_{i=1}^2 {\bf P}_i^2/(2m)+ V_i({\bf x}_i-\bar{{\bf x}}_i)$,
where ${\bf P}_i$ is the momentum of molecule $i$ with mass $m$, and
the potential generated by the optical lattice
$V_i(\bf{x}-\bar{\bf{x}}_i)$ describes an external confinement of
molecule $i$ about a local minimum $\bar{\bf{x}}_i$ with $1$D rms
width $z_0$. We assume isotropic traps that are approximately
harmonic near the trap minimum with a vibrational spacing
$\hbar\omega_{\rm
  osc}$.  Furthermore, we assume that the molecules can be prepared in
the motional ground state of each local potential using dissipative
electromagnetic pumping \cite{DeMille:05}.  It is convenient to
define the quantization axis $\hat{z}$ along the axis connecting the
two molecules, $\bar{\bf{x}}_2-\bar{\bf{x}}_1=\Delta z \hat{z}$ with
$\Delta z$ corresponding to a multiple of the lattice spacing.

The near field dipole-dipole interaction between two molecules
separated by ${\bf r}={\bf x}_1-{\bf x}_2$ is
\begin{equation}
  H_{\rm dd}=
  \frac{d^2}{r^3}\sum_{q=-1}^1 ((-1)^qD^{\dagger}_{1q}D_{2-q}-3D^{\dagger}_{10}D_{20}
  +h.c.).
  \label{dd}
\end{equation}
The dipole operator coupling the ground and first rotational states
of molecule $i$ is ${\bf D}^{\dagger}_i=\sum_{q=-1}^1 \ket{N=1,q}_i
{_i}\bra{N=0,0}\hat{e}^*_q$, and $d$ is the dimensionful dipole
moment.  

While the present situation of dipole-dipole coupling of
rotationally excited polar molecules is reminiscent of the
dipole-dipole interaction of between electronically excited atom
pairs \cite{GKB:02}, there are important differences. First, unlike
the atomic case where electronically excited states typically are
antitrapped by an optical lattice, here both ground and excited
rotational states are trapped by an essentially identical potential.
Hence motional decoherence due to spin dependent dipole-dipole
forces is strongly suppressed by the large vibrational energy
$\hbar\omega_{\ osc}$.  Second, scattering rates are drastically
reduced.  The decay rate at room temperature from
excited rotational states is $\sim 10^{-3}$\ Hz
\cite{Kotochigova:04} versus a comparable rate of MHz for excited
electronic states.

The ground subspace of each molecule is isomorphic to a spin $1/2$
particle.  Our goal is to obtain an effective spin-spin interaction
between two neighboring molecules.  Static spin-spin interactions
due to spin-rotation and dipole-dipole couplings do exist but are
very small in our model: $H_{\rm vdW}(r) = -(d^4/2Br^6)
\left[1+\left(\gamma/4B\right)^2\left(1+4{\bf
      S}_1\cdot{\bf S}_2/3-2S_1^zS_2^z \right) \right] $.  The first
term is the familiar van der Waals $1/r^6$ interaction, while the
spin dependent piece is strongly suppressed as $\gamma / 4B
\approx10^{-3} \ll 1$.  Therefore, we propose dynamical mixing with
dipole-dipole coupled excited states using a microwave field.

\begin{figure}[htb]
  \begin{center}
   \includegraphics[width=0.99\columnwidth]{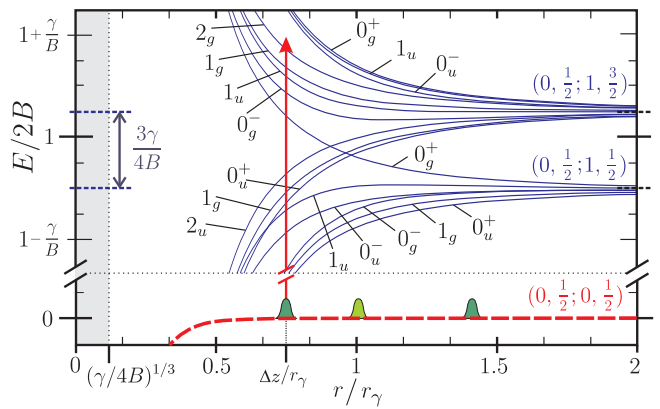}
    \caption{\label{fig:2}(Color online) Movre-Pichler potentials for
      a pair of molecules as a function of their separation $r$: The
      potentials $E(g_i(r))$ for the $4$ ground-state (dashed lines)
      and the potentials $E(\lambda(r))$ for the first $24$ excited
      states (solid lines). The symmetries $|Y|^\pm_\sigma$ of the
      corresponding excited manifolds are indicated, as are the
      asymptotic manifolds $(N_i,J_i;N_j,J_j)$.  The relative coordinate probability
      densities on a square lattice  are depicted on the ground state
      potential.
      }
  \end{center}
\end{figure}

The molecules are assumed trapped with a separation $\Delta z\sim
r_\gamma\equiv(2d^2/\gamma)^{1/3}$, where the dipole dipole
interaction is $d^2/r_\gamma^3=\gamma/2$.  In this regime the
rotation of the molecules is strongly coupled to the spin and the
excited states are described by Hunds case (c) states in analogy to
the dipole-dipole coupled excited electronic states of two atoms
with fine-structure.  The ground states are essentially spin
independent.  In the subspace of one rotational quantum
$(N_1+N_2=1)$, there are $24$ eigenstates of $H_{\rm in}$ which are
linear superpositions of two electron spin states and properly
symmetrized rotational states of the two molecules.  There are
several symmetries that reduce $H_{\rm in}$ to block diagonal form.
First, $H_{\rm dd}$, conserves the quantum number $Y=M_N+M_S$ where
$M_N=M_{N_1}+M_{N_2}$ and $M_S=M_{S_1}+M_{S_2}$ are the total
rotational and spin projections along the intermolecular axis.
Second, parity, defined as the interchange of the two molecules
followed by parity though the center of each molecule, is conserved.
The $\sigma=\pm 1$ eigenvalues of parity are conventionally denoted
$g(u)$ for {\it g}erade({\it u}ngerade). Finally, there is a
symmetry associated with reflection $R$ of all electronic and
rotational coordinates through a plane containing the intermolecular
axis.  For $|Y|>0$ all eigenstates are even under $R$ but for states
with zero angular momentum projection there are $\pm 1$ eigenstates
of $R$.  The $16$ distinct eigenvalues correspond to degenerate
subspaces labeled $|Y|_{\sigma}^{\pm}(J)$ with $J$ indicating the
quantum number in the $r\rightarrow \infty$ asymptotic manifold
$(N=0,J=1/2;N=1,J)$. Remarkably, the eigenvalues and eigenstates can
be computing analytically yielding the well known Movre-Pichler
potentials~\cite{Movre:77} plotted in Fig.~\ref{fig:2}.

\section{ENGINEERING SPIN-SPIN INTERACTIONS}

In order to induce strong dipole-dipole coupling we introduce a
microwave field $E({\bf x},t){\bf e}_F$ with a frequency $\omega_F$
tuned near resonance with the $N=0\rightarrow N=1$ transition.
Because the rotational states are spaced nonlinearly, this
transition is resolvable without coupling to higher rotational
states by multiphoton processes.  In the rotating wave
approximation, the molecule-field interaction is $H_{\rm
mf}=-\sum_{i=1}^2 (\hbar \Omega {\bf D}^{\dagger}_i\cdot {\bf
e}_Fe^{i({\bf k}_F\cdot {\bf
    x}_i-\omega_F t)}/2+h.c.)$, where the Rabi frequency is
$|\Omega|=d|E_0|/\hbar$. As the molecules are trapped by optical
wavelengths such that $k_F\Delta z \sim 10^{-5}$ the dipoles are
excited in phase only.

The effective Hamiltonian acting on the ground states is obtained in
second order perturbation theory as
\begin{equation}
  H_{\rm eff}(r)=\sum_{i,f}\sum_{\lambda(r)}\frac{\bra{g_f}H_{\rm mf}\ket{\lambda(r)}\bra{\lambda(r)}H_{\rm mf}\ket{g_i}}{\hbar\omega_F-E(\lambda(r))}\ket{g_f}\bra{g_i},
  \label{effHam1}
\end{equation}
where $\{{\ket{g_i},\ket{g_f}}\}$ are ground states with $N_1=N_2=0$
and $\{\ket{\lambda(r)}\}$ are excited eigenstates of $H_{\rm in}$
with $N_1+N_2=1$ and with excitation energies $\{E(\lambda(r))\}$.  The reduced
interaction in the subspace of the spin degrees of freedom is then
obtained by tracing over the motional degrees of freedom.  For
molecules trapped in the ground motional states of isotropic
harmonic wells with rms width $z_0$ the wave function is separable
in center of mass and relative coordinates with the relative
coordinate wavefunction
\[
  \psi_{\rm rel}(r,\theta)=\frac{1}{\pi^{3/4}(2z_0)^{3/2}} e^{-(r^2+\Delta z^2-2r\Delta z \cos \theta)/8 z_0^2},
\]
where $\cos \theta={\bf r}\cdot \hat{z}/r$.  The effective spin-spin
Hamiltonian is then $H_{\rm spin}=\langle H_{\rm
  eff}(r)\rangle_{\rm rel}$.

The Hamiltonian in Eq.~\eqref{effHam1} is guaranteed to yield some
entangling interaction for appropriate choice of field parameters
but it is desirable to have a systematic way to design a spin-spin
interaction. Fortunately, the model presented here possesses
sufficient structure to achieve this essentially analytically.  The
effective Hamiltonian on molecules $1$ and $2$ induced by a
microwave field is
\begin{equation}
  H_{\rm eff}(r)=\frac{\hbar|\Omega|}{8}\sum_{\alpha,\beta=0}^3 \sigma^{\alpha}_1A_{\alpha,\beta}(r)\sigma^{\beta}_2,
  \label{effHam2}
\end{equation}
where $\{\sigma^{\alpha}\}_{\alpha=0}^3\equiv \{{\bf
  1},\sigma^x,\sigma^y,\sigma^z\}$ and $A$ is a real symmetric tensor.
  See App.~\ref{appendix} for an explicit form of the matrix coefficients as  a function of
  field polarization and frequency.

Eq.~\eqref{effHam2} describes a generic permutation symmetric two
qubit Hamiltonian.  The components $A_{0,s}$ describe a pseudo
magnetic field which acts locally on each spin and the components
$A_{s,t}$ describe two qubit coupling.  The pseudo magnetic field is
zero if the microwave field is linearly polarized but a real
magnetic field could be used to tune local interactions and, given a
large enough gradient, could break the permutation invariance of
$H_{\rm spin}$.

For a given field polarization, tuning the frequency near an excited
state induces a particular spin pattern on the ground states.  These
patterns change as the frequency is tuned though multiple resonances
at a fixed intermolecular separation.  In Table~\ref{tab:2} it is
shown how to simulate the Ising and Heisenberg interactions in this
way.  Using several fields that are sufficiently separated in
frequency, the resulting effective interactions are additive
creating a {\it spin texture} on the ground states.  The anisotropic
spin model $H_{XYZ}=\lambda_x \sigma^x\sigma^x+\lambda_y
\sigma^y\sigma^y+\lambda_z\sigma^z\sigma^z$ can be simulated using
three fields: one polarized along $\hat{z}$ tuned to $0_u^+(3/2)$,
one polarized along $\hat{y}$ tuned to $0_g^-(3/2)$ and one
polarized along $\hat{y}$ tuned to $0_g^+(1/2)$.  The strengths
$\lambda_j$ can be tuned by adjusting the Rabi frequencies and
detunings of the three fields.  Using an external magnetic field and
six microwave fields with, for example, frequencies and
polarizations corresponding to the last six spin patterns in
Table~\ref{tab:2}, arbitrary permutation symmetric two qubit
interaction are possible.

The effective spin-spin interaction along a different intermolecular
axis $\hat{z}'$ can be obtained by a frame transformation in the spherical basis.
  Writing
$\hat{z}'=D^{1\dagger}(\beta_1,\beta_2,\beta_3).(0,1,0)^{T}$, where
$D^j$ is the spin-j Wigner rotation, the effective Hamiltonian along
$\hat{z}'$ in the original coordinate system is obtained by the
following replacements to the field polarization vector and spin operators:
$(\alpha_-,\alpha_0,\alpha_+)^{T}\rightarrow
D^{1\dagger}(\beta_1,\beta_2,\beta_3).(\alpha_{-},\alpha_0,\alpha_{+})^{T}$
and $\sigma^{\alpha}\rightarrow
D^{1/2}(\beta_1,\beta_2,\beta_3)\sigma^{\alpha}D^{1/2\dagger}(\beta_1,\beta_2,\beta_3)$.
For example, using a $\hat{z}$ polarized field tuned near
$0_u^+(3/2)$ and a field polarized in the $\hat{x}-\hat{y}$ plane
tuned near $1_u(3/2)$ creates a Heisenberg interaction between any
two molecules separated by ${\bf r}$ with arbitrary orientation in space.

\begin{table}
  \caption{\label{tab:2}  Some spin patterns that result from Eq.~\eqref{effHam2}.  The field polarization is given with respect to the intermolecular axis $\hat{z}$ and the frequency $\omega_F$ is chosen to be near resonant with the indicated excited state potential at the internuclear separation $\Delta z$.  The sign of the interaction will depend on whether the frequency is tuned above or below resonance.}
  \begin{ruledtabular}
    \begin{tabular}{ccc}
      \hline
      Polarization &\  Resonance &\  Spin pattern\\
      \hline
      $\hat{x}$ &\  $2_g$ &\  $\sigma^z\sigma^z$\\
      \hline
      $\hat{z}$ &\  $0_u^+$ &\  $\vec{\sigma}\cdot\vec{\sigma}$ \\
      \hline
      $\hat{z}$ &\  $0_g^-$ &\  $\sigma^x\sigma^x+\sigma^y\sigma^y-\sigma^z\sigma^z$ \\
      \hline
      $\hat{y}$ &\  $0_g^-$ &\  $\sigma^x\sigma^x-\sigma^y\sigma^y+\sigma^z\sigma^z$ \\
      \hline
      $\hat{y}$ &\  $0_g^+$ &\  $-\sigma^x\sigma^x+\sigma^y\sigma^y+\sigma^z\sigma^z$ \\
      \hline
      $(\hat{y}-\hat{x})/\sqrt{2}$ &\  $0_g^+$ &\  $-\sigma^x\sigma^y-\sigma^y\sigma^x+\sigma^z\sigma^z$ \\
      \hline
      $\cos\xi\hat{x}+\sin\xi\hat{z}$ &\  $1_g$ &\  $\lambda_1(\sigma^x\sigma^z+\sigma^z\sigma^x)+\lambda_2\sigma^z\sigma^z$\\
      &\ &\ $+\lambda_3(\sigma^x\sigma^x+\sigma^y\sigma^y)$ \\
      \hline
      $\cos\xi\hat{y}+\sin\xi\hat{z}$ &\  $1_g$ &\  $\lambda_1(\sigma^y\sigma^z+\sigma^z\sigma^y)+\lambda_2\sigma^z\sigma^z$\\
      &\ &\ $+\lambda_3(\sigma^x\sigma^x+\sigma^y\sigma^y)$ \\
    \end{tabular}
  \end{ruledtabular}
\end{table}

\section{APPLICATIONS}

We now show how to engineer the spin model ${\rm I}$. 
  Consider a system of trapped molecules in a square
lattice with site coordinates in the $\hat{z}-\hat{x}$ plane
$\{\bar{{\bf x}}_{i,j}\}=\{ib\hat{z}+jb\hat{x};\ i,j\in
[1,\ell]\bigcap\mathbb{Z}\}$.  Illuminate the system with a
microwave field with linear polarization ${\bf e}_F=\cos\zeta
\hat{y}+\sin\zeta\hat{x}$ and field frequency $\omega_F$ tuned such
that the peak of the relative coordinate wavefunction at $r=b$ is
near resonant with the $2_g$ potential but far detuned from other
excited states.  Then the dominate interaction between nearest
neighbor molecules is of Ising type along each axis and we realize
$H_{\rm spin}^{({\rm I})}$ with
$J=(\hbar|\Omega|)^2\langle1/8(\hbar\omega_F-2B-\gamma/2-d^2/r^3)\rangle_{\rm
  rel}$.  For realistic parameters, this coupling can range from $10-100$ kHz,
  with the strength constrained by the
  trap spacing $(J\ll\hbar\omega_{\rm osc})$.
    The relative strength of the interactions along $\hat{z}$ and
$\hat{x}$ can be changed by rotating the angle $\zeta$ of
polarization out of plane.  Interactions between more distant
neighbors are relatively weak because the far off resonant coupling
at larger $r$ cannot distinguish the spin dependence of excited
states.

\begin{figure}[htb]
  \begin{center}
    \includegraphics[width=.99\columnwidth]{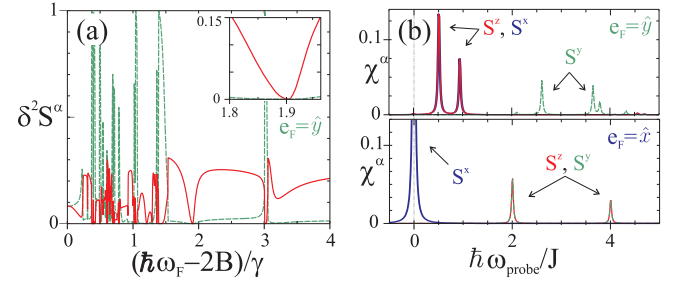}
    \caption{\label{fig:3}(Color online)  Design and verification of 
    noise protected ground states arising from a simulation of 
    $H_{\rm spin}^{({\rm I})}$.  The system is comprised of $9$ polar 
    molecules trapped in a $3\times 3$-square-lattice in the $\hat z-\hat
      x$ plane with lattice spacing $b=r_\gamma/\sqrt{2}$ driven with a 
      field of frequency $\omega_F$ and out of plane polarization angle
      $\zeta$.  (a)  Noise resilience of the ground states of the 
      resultant interaction $H_{\rm spin}$
      as a function of field frequency, quantified by the rms-magnetisations of the two ground-states,
      $\delta^2S^z=\delta^2S^x$ (solid-lines) and $\delta^2S^y$
      (dashed lines) for $\zeta=0$. The inset shows the protected region, when tuning
      near to the $2_g$ resonance $E(2_g)\approx 1.9\gamma$ which realizes
      the model $H_{\rm spin}^{({\rm I})}$. (b)  
      Absorption-spectroscopy of ground states $\chi^\alpha(\omega_{\rm probe})$ for two
      spin-textures obtained by tuning $\omega_F$ near the $2_g$
      resonance, $\delta/\gamma=1.88$, with $b=r_\gamma/\sqrt{2}$.  For
      $\zeta=0$ the spectrum is gapped by $J/2$, which is a signature of
      a protected qubit (top), while for $\zeta=\pi/2$ the
      excitations are gap-less spin-waves (bottom). The peak at
      $\omega_{\rm probe}=0$ is of order unity. The spectroscopic
      coupling component $\alpha=x,y,z$ (solid,dashed,dotted lines)
      are indicated.}

  \end{center}
\end{figure}

The authors of Ref. \cite{Duocot:05} show that the ideal spin model
${\rm I}$ (for $\zeta\neq \pm\pi/2$) has a $2$-fold
degenerate ground subspace, which is gapped with weak size
dependence for $\cos\zeta=1$.  The the two ground-states, which we 
denote, $\ket{0}_L$
and $\ket{1}_L$, have zero local magnetizations
$\langle\sigma^\alpha_{i,j}\rangle_L$.  Our implementation is not
ideal because there are residual longer range interactions along
several directions in the plane as well as off resonant couplings to
excited state potentials yielding unwanted spin patterns. We note,
however, that all the effective spin Hamiltonians described in
Eq.~\eqref{effHam2} obtained using fields with linear polarization
involve sums of products of Pauli operators and hence are invariant
under time-reversal. For $\ell$ odd, the degeneracy present in the
ground state of $H_{\rm spin}^{({\rm I})}$ is of Kramers' type and
imperfect implementation will not break this degeneracy though it
may decrease the energy gap.

We have numerically computed the effective interaction on a
$\ell^2=3\times3$ square lattice with spacings $b=r_\gamma/\sqrt{2}$
and we take the localization to the point dipole limit.  In
Fig.~\ref{fig:3}(a) we plot the $\alpha=x,y,z$-components of the rms
magnetization for the ground subspace, $\delta^2S^\alpha\equiv
\sum_{ij}\sum_{G'G}|_L\bra{G'}\sigma^\alpha_{i,j}\ket{G}_L|^2/2\ell^2$,
as a function of the detuning $\omega_F-2B/\hbar$ for polarization
angle $\zeta=0$. This allows for computing logical qubit errors due to
quasi-static noise.  Near the bare resonance $\hbar\omega_F-2B=\gamma/2$
the system show multiple long-range resonances as all the sites
couple near-resonantly at coupling strength $\propto 1/b^3$. The
last of these long-range resonance appears at
$\hbar\omega_F-2B\approx1.36\gamma$ for the interaction between next
nearest neighbor sites with spacings of $\sqrt{2}b$. The
$2_g$-resonance lies at $\hbar\omega_F-2B\approx1.9\gamma$ for
nearest neighbor sites and shows the remarkable feature of no
magnetization on any site in any space-direction $\alpha$ withing
the ground-state manifold (see inset).  The resulting immunity of
the system to local noise can be probed by applying an homogeneous
$B$-field of frequency $\omega_{\rm probe}$ polarized in the
direction $\alpha=x,y,z$.  The corresponding absorption spectrum for
an arbitrary code state $\ket{\psi}_L$ is, 
$\chi^\alpha(\omega_{\rm
  probe})\equiv-\hbar\Gamma\Im[{_L\bra{\psi}}S^\alpha(\hbar\omega_{\rm
  probe}-H_{\rm spin}+i\hbar\Gamma)^{-1}S^\alpha{\ket{\psi}_L}]$ where 
  $S^\alpha=\sum_{ij}\sigma^\alpha_{i,j}/\ell^2$ and
  $\Gamma$ is an effective linewidth.  This quantity is plotted in
Fig.~{\ref{fig:3}(b)} for two different
spin-textures obtained for the same field frequency $\omega_F=1.88\gamma$ but 
different polarizations and were we 
set $\Gamma=0.1J$.
For polarization
$\zeta=0$ (see top inset) one realizes the protected qubit, whose
spectrum is gapped by $J/2$. For polarization along the $\hat
x$-direction $\zeta=\pi/2$ (see bottom inset) the ground-subspace is
given by a set of $\ell$ quantum-Ising stripes along $z$, whose
spectrum is ungapped with a large peak at $\omega_{\rm probe}=0$ in 
response to coupling with a $B$ field polarized along $\alpha=x$.

\begin{figure}
  \begin{center}
    \includegraphics[width=.59\columnwidth]{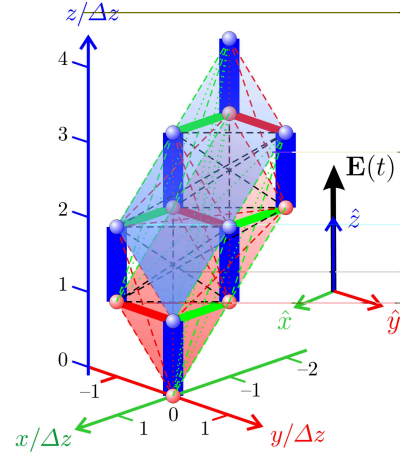}
    \caption{\label{fig:4}(Color online)  Implementation of spin model $H_{\rm spin}^{({\rm II})}$.
    Shown is the spatial configuration of $12$ polar molecules trapped by two parallel triangular 
    lattices (indicated by shaded planes) with separation normal to the plane of $\Delta z/\sqrt{3}$ and 
    in plane relative lattice shift of $\Delta z\sqrt{2/3}$.  Nearest neighbors are separated by $b=\Delta z$ and next nearest neighbor couplings are at $\sqrt{2}b$.  The graph vertices represent spins and the edges correspond to pairwise spin couplings.  The edge color indicates the nature of the dominant 
   pairwise coupling for that edge (blue$=\sigma^z\sigma^z$, red$=
   \sigma^y\sigma^y$, green$=\sigma^x\sigma^x$, black$=$``other").   For nearest neighbor couplings, the edge width indicates the relative strength of the absolute value of the coupling.
   For this implementation, the nearest neighbor separation is $b=r_{\gamma}$. 
   Three fields all polarized along $\hat{z}$ were used to generate the effective spin-spin interaction with frequencies and intensities optimized to approximate the ideal model $H_{\rm spin}^{({\rm II})}$.  The field detunings at the nearest neighbor spacing are:  $ \hbar\omega_1-E(1_g(1/2))=-0.05 \gamma/2, \hbar\omega_2-E(0_g^-(1/2))=0.05 \gamma/2,\hbar\omega_3-E(2_g(3/2))=0.10 \gamma/2$ and the amplitudes are $|\Omega_1|=4|\Omega_2|=|\Omega_3|=0.01 \gamma/\hbar$.  For $\gamma=40 {\rm MHz}$ this generates effective coupling strengths $J_z=-100 {\rm kHz}$ and $J_{\perp}=- 0.4 J_z$.  The magnitude of residual nearest neighbor couplings are less than $0.04|J_z|$ along $x$ and $y$-links and less than $0.003|J_z|$ along $z$-links.  The size of longer range couplings $J_{\rm lr}$ are indicated by edge line style (dashed: $|J_{\rm lr}|<0.01 |J_z|$, dotted:  $|J_{\rm lr}|<10^{-3} |J_z|$).    Treating pairs of spins on $z$-links as a single effective spin in the low energy sector, the model approximates Kitaev's $4$-local Hamiltonian \cite{DKL:03} on a square grid (shown here are one palquette on the square lattice and a neighbor plaquette on the dual lattice) with an effective coupling strength $J_{\rm eff}=-(J_{\perp}/J_z)^4|J_z|/16\approx 167 {\rm Hz}$.}
  \end{center}
\end{figure}

Spin model ${\rm II}$ is likewise obtained
using this mechanism. Consider a system of four molecules connected
by three length $b$ edges forming an orthogonal triad in space.  There are
several different microwave field configurations that can be used to 
realize the interaction $H_{\rm spin}^{({\rm II})}$ along the links.  One choice is to 
use two microwave fields polarized along
$\hat{z}$, one tuned near resonance with a $1_g$ potential and one
near a $1_u$ potential.  When the detunings and Rabi frequencies are
chosen so that $\langle|\Omega_{1_g}|
C(1_g,3,3)-|\Omega_{1_u}|C(1_u,1,1)\rangle_{\rm rel}=0$ then the
resultant spin pattern is Eq. \eqref{Kit} with $J_{\perp}=-\hbar
\langle |\Omega_{1_g}|C(1_g,3,3)\rangle_{\rm rel}/4$ and $J_z=\hbar|
\langle |\Omega_{1_g}|C(1_g,2,2)\rangle_{\rm rel}/4$. The ratio
$|J_{\perp}|/|J_z|$ can be tuned either by changing the lattice
spacing or by using a third microwave field polarized along
$\hat{z}$ and tuned near the $2_g$ potential, in which case
$J_{\perp}\rightarrow J_{\perp}+\hbar \langle
|\Omega_{2_g}|C(2_g)\rangle_{\rm rel}/8$. A bipartite lattice
composed of such triads with equally spaced nearest neighbors can be
built using two planes of stacked triangular lattices.
  Such a lattice could be 
designed using bichromatic trapping lasers in two spatial dimensions and 
a suitably modulated lattice in the third dimension normal to both planes.
 A realization of model {\rm II} using 
 a different set of $3$ microwave fields is shown in Fig. \ref{fig:4}.
The obtained interaction is close to ideal with small
residual coupling to next nearest neighbors as in model ${\rm I}$.

\section{CONCLUSIONS}

We have shown how to engineer pairwise spin-1/2 (or qubit)
interactions between polar molecules with designable range and
spatial anisotropy. The couplings are strong relative to decoherence
rates, and the simulation does not require complicated control
sequences to build effective time averaged Hamiltonians,
spin-dependent lattices or local addressability. Thus polar
molecules in optical lattices are a good candidate to provide a
laboratory realization of exotic states of matter. We envision that
engineering of these new materials might eventually provide the
basis of a new type of quantum optics, where systematic tools are
developed which allow the controlled preparation and manipulation of
excitations such as anyons, with applications ranging from
fundamental questions in condensed matter physics to quantum
computing.

\section{Acknowledgements}
A.~M. thanks W.~Ernst, and P.Z. thanks T. Calarco, L. Faoro,
M.~Lukin, D.~Petrov for helpful discussions. This work was supported
by the Austrian Science Foundation, the European Union, and the
Institute for Quantum Information.


\appendix
\section{Effective Interactions}
\label{appendix}

The effective spin-spin interaction Eq. \eqref{effHam2} between polar
molecules depends both on the frequency $\omega_F$ and polarization
${\bf e}_F=\alpha_{-}\hat{e}_{-1}+\alpha_{0}\hat{e}_{0}+\alpha_{+}\hat{e}_{1}$,
$(\hat{e}_0\equiv\hat{z})$ of the field.  The explicit form for the coupling
coefficients is:
\[
  \begin{array}{lll}
    A_{1,1}&=& |\alpha_0|^2[C(0_g^-,1,2)-C(0_u^+,1,2)]\\
    & &+(|\alpha_-|^2+|\alpha_+|^2)[C(1_g,3,3)-C(1_u,1,1)]\\
    & &+\Re[\alpha_+^*\alpha_-][C(0_g^-,2,1)-C(0_g^+,2,1)],\\
    A_{2,2}&=&A_{1,1}-2\Re[\alpha_+^*\alpha_-][C(0_g^-,2,1)-C(0_g^+,2,1)],\\
    A_{3,3}&=& |\alpha_0|^2[2C(1_g,2,2)-C(0_g^-,1,2)-C(0_u^+,1,2)]\\
    & &+(|\alpha_+|^2+|\alpha_-|^2) [C(2_g)+C(0_g^+,2,1)/2\\
    & &+C(0_g^-,2,1)/2-C(1_u,1,1)-C(1_g,3,3)],\\
    A_{1,2}&=&\Im [\alpha_+^*\alpha_-](C(0_g^-,2,1)-C(0_g^+,2,1)),\\
    A_{1,3}&=&\Re [\alpha_+^*\alpha_0-\alpha_0^*\alpha_-]C(1_g,2,3),\\
    A_{2,3}&=&\Im [\alpha_+^*\alpha_0-\alpha_0^*\alpha_-]C(1_g,2,3),\\
    A_{0,1}&=&\Re [\alpha_+^*\alpha_0+\alpha_0^*\alpha_-]C(1_g,2,3),\\
    A_{0,2}&=&\Im [\alpha_+^*\alpha_0+\alpha_0^*\alpha_-]C(1_g,2,3),\\
    A_{0,3}&=& (|\alpha_+|^2-|\alpha_-|^2)[C(2_g)-C(0_g^+,2,1)/2\\
    & &-C(0_g^-,2,1)/2].\;
  \end{array}
\]
The component $A_{0,0}$ weights a scalar energy shift which we
ignore. The coefficients $C(|Y|_{\sigma}^{\pm})$ quantify
coupling to excited states with different symmetries and are given
by
\[
\begin{array}{lll}
  C(0_{\sigma}^m,j,k)&=&K_j(0_{\sigma}^m)^2 s(0_{\sigma}^m(3/2))+K_k(0_{\sigma}^m)^2 s(0_{\sigma}^m(1/2)),\\
  C(1_{\sigma},j,k)&=&\sum_{a=1}^4 K_j^a(^{a}1_{\sigma}(3/2))K_k^a(^{a}1_{\sigma}(3/2)) s(^{a}1_{\sigma}(3/2))\\
  &+&\sum_{b=1}^2 K_j^b(^{b}1_{\sigma}(1/2))K_k^b(^{b}1_{\sigma}(1/2)) s(^{b}1_{\sigma}(1/2))\\
  C(2_g)&=&s(2_g(3/2)).
  \end{array}
\]
Here the energy dependent terms
$s(|Y|_{\sigma}^{\pm}(J))=\hbar|\Omega|/[\hbar\omega_F-E(|Y|_{\sigma}^{\pm}(J))]$
quantify the amplitude in the excited states. The energies
$E(|Y|_{\sigma}^{\pm}(J)))$ correspond to eigenvalues of $H_{\rm
int}$ and the sets $\{K_j(0^m_{\sigma}(J))\}_{j=1}^2$ and
$\{K_j^a(^{a}1_{\sigma}(J))\}_{j=1}^3$ are coefficients of the
eigenvectors for $|Y|=0,1$. For $Y=0$, the energies are
 $E(0_{\sigma}^+(1\pm 1/2))=2B+ \gamma\big[\sigma 3d^2/2\gamma r^3-1/4\pm
\sqrt{(\sigma d^2/2\gamma r^3+1/4)^2+1/2}\big]$, and
$E(0_{\sigma}^-(1\pm 1/2))=2B+\gamma \big[-\sigma d^2/2\gamma
r^3-1/4\pm \sqrt{(-\sigma 3d^2/2\gamma r^3+1/4)^2+1/2}\big]$,
The eigenvector components are
$K_1(0^m_{\sigma})=\cos(\gamma_{O_{\sigma}^m}/2)$ and
$K_2(0^m_{\sigma})=\sin(\gamma_{O_{\sigma}^m}/2)$ where the angles
satisfy $\tan(\gamma_{O_{\sigma}^+})=\sqrt{2}/(1/2+\sigma d^2/\gamma
r^3)$, and $\tan(\gamma_{O_{\sigma}^-})=\sqrt{2}/(1/2-\sigma
3d^2/\gamma r^3)$.  For $Y=\pm 1$ the eigenvectors and doubly
degenerate eigenvalues are obtained by diagonalizing the $3\times 3$
matrices:
\[
2B{\bf 1}_3+\frac{\gamma}{2}
\begin{pmatrix}
    -\sigma\frac{2d^2}{\gamma r^3}  & \pm 1 & \mp 1 \\
   \pm 1   & -\sigma \frac{4d^2}{\gamma r^3} & 1 \\
    \mp 1  & 1 & \sigma\frac{2d^2}{\gamma r^3}
\end{pmatrix}.
\]
For $Y=\pm 2$, the eigenvalues are doubly degenerate with energies
$E(2_{\sigma}(3/2))=2B+\gamma/2+\sigma d^2/r^3$.

A caveat is that we do not have point dipoles but rather wavepackets
with spatial distributions parallel and perpendicular to the
intermolecular axis $\hat{z}$.  Components of intermolecular
separations orthogonal to $\hat{z}$ will couple to states with
different symmetry and an exact treatment would require averaging
over the angular distrubution with the appropriate frame
transformation.  However, we argue that in our regime this finite
size affect is negligible.  The relative magnitude can be estimated
by the ratio of the marginal relative coordinate probability
distributions perpendicular and parallel to $\hat{z}$.  Defining
$p_{\bot}(r)=\int d\Omega \sin^2\theta r^2 |\psi_{\rm
rel}(r,\theta)|^2$ and $p_{\|}(r)=\int d\Omega \cos^2\theta r^2
|\psi_{\rm rel}(r,\theta)|^2$, the peak of the distributions is at
$r=\Delta z$ where for $z_0/\Delta z\ll 1$, the relative amount of
unwanted couplings is $p_{\bot}(\Delta z)/p_{\|}(\Delta z)\sim
4(z_0/\Delta z)^2$.  For molecular wavepacket localization $2\pi
z_0/\lambda_{\rm trap}=0.1$, the ratio is
$p_{\bot}( \lambda_{\rm trap})/p_{\|}(\lambda_{\rm trap})\approx 10^{-3}$,
hence it is warranted to compute the couplings as if the entire
weight of the wavefunction were parallel to $\hat{z}$.

\end{document}